# Causal Reasoning in Graphical Time Series Models

**Michael Eichler**
Department of Quantitative Economics
University of Maastricht, NL
m.eichler@ke.unimaas.nl

**Vanessa Didelez**
Department of Statistical Science
University College London, UK
vanessa@stats.ucl.ac.uk

**Abstract**

We propose a definition of causality for time series in terms of the effect of an intervention in one component of a multivariate time series on another component at some later point in time. Conditions for identifiability, comparable to the back–door and front–door criteria, are presented and can also be verified graphically. Computation of the causal effect is derived and illustrated for the linear case.

## 1 INTRODUCTION

In the time series literature, the notion of causality is closely linked to the seminal ideas of Granger (1969). Roughly speaking, he proposed to call a time series noncausal for another series if the past of the former does not predict the future of the latter given other relevant information about the past. One might infer from this that if a time series is not noncausal in Granger's sense then it is *potentially* causal, where the emphasis is on 'potentially' due to the well–known fact that 'association is not causation.'

Here, we propose a definition and identifiability criteria for when a time series is *actually* causal for another series. Our approach is based on the notion of interventions, external changes to the system, which we regard as being at the core of causal reasoning — typically one has a future (or past) intervention in mind when the term cause is used. Hence, we consider the effect of an intervention in one component of the time series, at a given point in time, on another (or the same) component at a later point in time.

We make use of a special graphical representation of the dependence structure within multivariate time series in order to capture the dynamic nature of the dependencies (Eichler, 2005, 2006, 2007). This is based on conditional independencies between the past and future of time series, leading us back to the ideas of Granger. More specifically, noncausality in Granger's sense will be represented by the absence of an arrow in the graph as detailed in Section 4.3. We can then formulate graphical criteria that allow identification of the effect of an intervention in a way resembling the back– and front–door criteria (Pearl, 1995).

The outline of the paper is as follows. We set the scene in Section 2, formalising in particular our understanding of causality as effect of an intervention. Section 3 gives the time series version of the back–door criterion. The main results are presented in Secion 4, where we address the graphical verification of identifiability.

## 2 PRELIMINARIES

Throughout we consider a multivariate stationary time series $X = \{X(t), t \in \mathbb{Z}\}$ with $X(t) = (X_1(t), \ldots, X_d(t))'$. Let $V$ be the index set $\{1, \ldots, d\}$. For any $A \subseteq V$ we define $X_A = \{X_A(t)\}$ as the multivariate subprocess with components $X_a$, $a \in A$. Furthermore, $\overline{X}_A(t)$ denotes the history of $X_A$ up to and including $t$, i.e. the set $\{X_A(s), s \leq t\}$. Throughout we assume that $X$ is a stationary, mixing stochastic process and that its conditional distributions, denoted by $\mathbb{P}(X(t+1)|\overline{X}(t))$, have regular almost surely absolutely continuous versions.

### 2.1 INTERVENTIONS

The key to our definition of causal effect are so–called intervention indicators (Pearl, 1993; Spirtes et al., 2000; Lauritzen, 2001; Dawid, 2002), denoted by $\sigma_a(t)$ where $\sigma$ stands for 'strategy', indicating whether or not we intervene in the system and manipulate $X_a(t)$, $a \in V$, or whether we leave 'nature' free reign (cf. also Didelez et al., 2006). More specifically, the role of $\sigma$ is defined as follows.

**Definition 2.1 (Regimes)** Let $X$ be a multivari-

ate stationary time series. The intervention indicator $\sigma_a(t)$, $a \in V$, takes values in $\{\emptyset, s \in \mathcal{S}\}$ with the following interpretations.

(i) *Idle Regime:* When $\sigma_a(t) = \emptyset$ we let $X_a(t)$ arise naturally without intervention. The conditional distributions of $X_a(t)$ given $\sigma_a(t)$ and possibly other components are not necessarily known. We also call this the observational regime.

(ii) *Atomic interventions:* Here $\mathcal{S} = \mathcal{X}$, the domain of $X_a(t)$, such that $\sigma_a(t) = x^*$ means we intervene and force $X_a(t)$ to assume the value $x^*$. In particular this means that

$$\mathbb{P}(X_a(t) = x | \overline{X}_V(t-1); \sigma_a(t) = x^*)$$
$$= \mathbb{P}(X_a(t) = x | \sigma_a(t) = x^*) = \delta_{\{x^*\}}(x),$$

where $\delta_D(x)$ is one if $x \in D$ and zero otherwise.

(iii) *Conditional intervention:* Here $\mathcal{S}$ consists of functions $g(\overline{x}_C(t-1)) \in \mathcal{X}$, $C \subset V$, such that $\sigma_a(t) = g$ means $X_a(t)$ is forced to take on a value that depends on past observations of $\overline{X}_C(t-1)$, i.e.

$$\mathbb{P}(X_a(t) = x | \overline{X}_V(t-1); \sigma_a(t) = g)$$
$$= \mathbb{P}(X_a(t) = x | \overline{X}_C(t-1); \sigma_a(t) = g)$$
$$= \delta_{\{g(\overline{X}_C(t-1))\}}(x).$$

(iv) *Random intervention:* Here $\mathcal{S}$ consists of distributions meaning that $X_a(t)$ is forced to arise from such a distribution, i.e. the conditional distribution $\mathbb{P}(X_a(t) | \overline{X}_V(t-1); \sigma_a(t) = s)$ is *known* and possibly a function of $\overline{X}_C(t-1)$, $C \subseteq V$.

**Remark 2.2 (Non–randomness of $\sigma$)** The intervention indicator $\sigma_a(t)$ is not a random variable. It is a decision variable and indicates different situations under which data could be generated or observed and hence indexes the corresponding distributions.

In the following we use the symbol $\perp\!\!\!\perp$ and properties of conditional independence as presented by Dawid (1979). When this symbol is applied to the intervention indicator, e.g. if we write $X_b(t+h) \perp\!\!\!\perp \sigma_a(t)$, it means that the distribution of $X_b(t+h)$ is the same under any regime considered (Dawid, 2002). Similarly we write $\mathbb{E}_\emptyset$ and $\mathbb{E}_{\sigma_a(t)=s}$ to distinguish between expectations with respect to the idle regime or under a specific intervention $s$ (the shorthand $\mathbb{E}_s$ is used when it is clear from the context what variable is intervened in). In contrast $\mathbb{E}^\mathcal{F} X$ denotes conditional expectation given $\mathcal{F}$ (e.g. Kallenberg, 2001).

We will make the following assumptions about how an intervention in $X_a(t)$ affects the remaining system. These assumptions have to be checked in any given application as to whether the particular intervention of interest satisfies them, and will typically also depend on what components $X_1, \ldots, X_d$ are included in the multivariate time series.

**Assumption 2.3 (Intervention)** Let $X$ be a multivariate stationary time series. An intervention in $X_a(t)$ is assumed to have the following properties.

(i) $(\overline{X}_V(t-1), X_{V \setminus \{a\}}(t)) \perp\!\!\!\perp \sigma_a(t)$;

(ii) $\{X_V(t+j), j \in \mathbb{N}\} \perp\!\!\!\perp \sigma_a(t) | \overline{X}_V(t)$.

Part (i) ensures that whether or not we intervene in $X_a(t)$ is not informative or informed by any of the earlier variables or the remaining contemporaneous variables. In particular this excludes instantaneous causality, as is justified when the variables $X_V(t)$ truely arise at the same time.

The second assumption ensures that future variables are not associated with the intervention other than through past variables. The two together are meant to reflect our concept of an intervention being an isolated exogenous change of the system.

**Remark 2.4 (Multiple interventions)** Interventions and their indicators can be extended to the case of more than one variable, e.g. $\sigma_A(t)$, $A \subset V$, as well as to more than one point in time, e.g. $\sigma_a(k)$, $k \in K \subset \mathbb{Z}$. In that case each $\sigma_a(k)$, $a \in A$, $k \in K$, is assumed to satisfy the above conditions individually. In the following we mostly only consider the case of one component and one point in time, though multiple interventions are implicitly needed for the front–door criterion in Section 4.5.

An obvious question to address is, why we should believe in Assumptions 2.3 for a given data situation. In particular in cases of 'confounding' it is well known that whether or not we intervene is informative for variables other than only the one that is manipulated because confounding typically means that the data (in the observational regime) might exhibit some kind of association that would not be present under an experimental setting. It is, in fact, the presence of such potential confounding that motivates the methods proposed in the following sections. The idea is to include in our considerations unobservable variables or processes such that we can reasonably believe, based on substantive background knowledge, that Assumptions 2.3 are satisfied. Then we give sufficient criteria that guarantee that the causal effect can be identified based on the observable processes alone. In the following we regard the whole multivariate time series $X = X_V = (X_1, \ldots, X_d)$ as the system that includes all components judged relevant but not necessarily observable, whereas $X_S$, $S \subseteq V$, will denote a reduced subprocess.

## 2.2 CAUSAL EFFECT

In general, the causal effect of an intervention can be any function of the post–intervention distribution of $\{X_V(t+j)|j \in \mathbb{N}\}$ given $\sigma_a(t) = s$. We define below the causal effect of intervening in $X_a(t)$ on $X_b(t+h)$ as the average of its post intervention distribution.

**Definition 2.5 (Average causal effect)** The *average causal effect (ACE)* of $X_a(t)$ on $X_b(t+h)$, $a, b \in V, h > 0$ following strategy $s$ is given by

$$\text{ACE}_s = \mathbb{E}_{\sigma_a(t)=s} X_b(t+h).$$

As $\mathbb{E}_\emptyset X_b(t+h) = 0$, the $\text{ACE}_s$ can be regarded as the average difference between no intervention and strategy $s$. Also, we can compare different strategies e.g. by considering $\text{ACE}_{s_1} - \text{ACE}_{s_2}$

Even though we focus on the ACE, the results presented in this paper hold more generally for $\mathbb{E}_{\sigma_a(t)} f(X_b(t+h))$ for any measurable function $f$ and thus for the post–intervention distribution $\mathbb{P}(X_b(t+h)|\sigma_a(t) = s)$ itself.

## 3 IDENTIFICATION OF ACE

A priori there is no reason why data that is *not* collected under the regime of interest should allow estimation of the ACE. By identifiability we mean the possibility to express the ACE in terms of quantities that are known or estimable under the observational regime.

The criterion presented below ensures such an identification of the causal effect. It is called 'back-door' criterion due to the graphical way of checking it, which will be presented in Section 4, where we will also briefly sketch another criterion called 'front-door' criterion.

**Theorem 3.1 (Back-door criterion)** *Let $a, b \in S \subseteq V$. Suppose that Assumptions 2.3 hold and $X_b(t+h) \perp\!\!\!\perp \sigma_a(t) \mid \overline{X}_S(t)$ for all $h \in \mathbb{N}$. If a conditional intervention $s$ is considered, as in Definition 2.1(iii, iv), then we also assume that the conditioning components are contained in $S$, i.e. $C \subset S$.
Then $S$ identifies the effect of $X_a(t)$ on $X_b(t+h)$ for all $h \in \mathbb{N}$, and the $ACE_s$ is given by*

$$\mathbb{E}_s X_b(t+h) = \mathbb{E}_\emptyset \mathbb{E}_s^{\overline{X}_a(t-1), \overline{X}_{S\setminus a}(t)} \mathbb{E}_\emptyset^{\overline{X}_S(t)} X_b(t+h). \quad (1)$$

**Proof:** As we assume $X_b(t+h) \perp\!\!\!\perp \sigma_a(t)|\overline{X}_S(t)$ it follows that

$$\mathbb{E}_s^{\overline{X}_S(t)} X_b(t+h) = \mathbb{E}_\emptyset^{\overline{X}_S(t)} X_b(t+h)$$

yielding the term on the very right of (1). The expectation in the middle of (1) is known as $C \subset S$. Furthermore, from Asumptions 2.3 and the properties of conditional independence, we have $\overline{X}_S(t-1), X_{S\setminus\{a\}}(t) \perp\!\!\!\perp \sigma_a(t)$, which yields $\mathbb{E}_s Z = \mathbb{E}_\emptyset Z$ for all $\sigma\{\overline{X}_S(t-1), X_{S\setminus\{a\}}(t)\}$–measurable random variables $Z$, which yields the first expectation of (1) and concludes the proof.

In (1) we can estimate $\mathbb{E}_\emptyset^{\overline{X}_S(t)} X_b(t+h)$ from observational data, while $\mathbb{E}_s^{\overline{X}_a(t-1), \overline{X}_{S\setminus a}(t)}$ is the expectation w.r.t. the intervention which is fully known. The outer expectation is again observational. Hence, provided that $X_S$ has been observed, we can use the above to estimate the causal effect. Dawid (2002) calls such a set $S$ 'sufficient covariates' or 'de–confounder'. Note that under Assumptions 2.3, $V$ always has to identify the causal effect due to condition (ii). In this sense we can say that the whole system $V$ contains all 'relevant' variables or components.

**Example 3.2** Let $X$ be a purely nondeterministic stationary Gaussian process with spectral matrix $f(\lambda)$, $\lambda \in [-\pi, \pi]$ such that the eigenvalues of $f(\lambda)$ are bounded and bounded away from zero uniformly for all $\lambda \in [-\pi, \pi]$. Furthermore, suppose that we are interested in the average causal effect of setting $X_a(t)$ to the value $x^*$ and that the effect is identified by the variables in $S$. By the assumptions on the spectral matrix, the subprocess $X_S$ has a mean-square convergent autoregressive representation

$$X_S(t) = \sum_{j=1}^\infty \Phi(j) X_S(t-j) + \varepsilon_S(t), \quad (2)$$

where $\varepsilon(t), t \in \mathbb{Z}$, are independent and identically normally distributed with mean zero and non-singular covariance matrix $\Sigma$. Moreover, the best $h$-step predictor $\mathbb{E}_\emptyset^{\overline{X}_S(t)} X_b(t+h)$ is equal to the best linear $h$-step predictor, that is,

$$\mathbb{E}_\emptyset^{\overline{X}_S(t)}(X_b(t+h)) = \sum_{j\in\mathbb{N}} \sum_{s\in S} \Phi_{bs}^{(h)}(j) X_s(t-j+1). \quad (3)$$

It follows from Theorem 3.1 that the causal effect of an intervention $s$ setting $X_a(t)$ to $x^*$ is given by

$$\mathbb{E}_s(X_b(t+h)) = \Phi_{ba}^{(h)}(1) x^*,$$

that is, the ACE is identical to the best linear predictor given a single observation $X_a(t) = x^*$. The coefficient $\Phi_{ba}^{(h)}(1)$ of the multi-step predictor can be computed recursively from the coefficients of the autoregressive representation in (2) using the relations

$$\Phi_{ba}^{(h)}(1) = \sum_{j=1}^{h-1} \sum_{s\in S} \Phi_{bs}^{(1)}(j) \Phi_{sa}^{(h-j)}(1) + \Phi_{ba}^{(1)}(h)$$

where $\Phi^{(1)}(j) = \Phi(j)$ (Box et al., 1994, Section 5.3).

## 4 GRAPHICAL CRITERIA

We suggest a graphical representation of the dependencies which allows us to link Granger–(non)causality to intervention causality and to read off the graphs whether the back–door criterion is met.

### 4.1 GRANGER–NONCAUSALITY

The following notion of strong Granger–noncausality (e.g. Florens and Mouchart, 1982) forms the base of our graphical time series models.

**Definition 4.1 (Granger–noncausality)** Let $X = X_V$ be a stationary multivariate time series. Let $A$ and $B$ be disjoint subsets of $V$ and let $X_A$, $X_B$ be the corresponding subprocesses of $X$.

(i) Then $X_A$ is *(strongly) Granger–noncausal for $X_B$ up to horizon $h$*, $h \in \mathbb{N}$, with respect to the process $X_V$ if

$$X_B(t+k) \perp\!\!\!\perp \overline{X}_A(t) | \overline{X}_{V \setminus A}(t)$$

for all $k = 1, \ldots, h$ and $t \in \mathbb{Z}$. If the above holds only for $h = 1$ we simply say that $X_A$ is (strongly) Granger–noncausal for $X_B$ with respect to $X_V$, and this will be denoted by $X_A \not\to X_B[X_V]$. If the above holds for all $h \in \mathbb{N}$ we say that $X_A$ is (strongly) Granger–noncausal for $X_B$ at *all horizons*, and this will be denoted by $X_A \overset{\infty}{\not\to} X_B[X_V]$.

(ii) The processes $X_A$ and $X_B$ are *contemporaneously independent* with respect to the process $X_V$ if

$$X_A(t+1) \perp\!\!\!\perp X_B(t+1) | \overline{X}_V(t)$$

for all $t \in \mathbb{Z}$. Contemporaneous independence will be denoted by $X_A \not\sim X_B[X_V]$.

Strong Granger–noncausality means that the past of $X_B$ up to time $t$ does not help to predict $X_A$ at the next point in time $t + 1$ given information about the past of all the remaining components (including $X_A$'s past). In contrast, strong Granger–noncausality at *all horizons* implies that this holds for any time in the future. The latter is more restrictive and if not stated otherwise we will only deal with strong Granger–noncausality at horizon $h = 1$. Both properties are 'strong' because they are phrased in terms of stochastic independence, not correlation; but for ease of notation we usually drop the 'strong'. A reason for the possible presence of contemporaneous dependence are hidden variables or processes not contained in $X_1, \ldots, X_d$ which might induce dependencies that cannot be explained by the observed past. This will play an important role when it comes to causal inference.

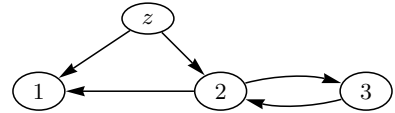

Figure 1: Mixed graph associated with the processes $X$ and $Z$ in Example 4.2.

We represent Granger–noncausality graphically by identifying the components of $X$ with the nodes of a graph, where the absence of a directed edge $a \to b$ means $X_a \not\to X_b [X_V]$ and the absence of a undirected (dashed) edge $a$ --- $b$ stands for $X_a \not\sim X_b [X_V]$.

We note that more detailed representations of the dynamic dependencies among the components of $X$ could be obtained by treating every variable at every point in time separately (e.g. Dahlhaus and Eichler, 2003; Moneta and Spirtes, 2005). However, this can lead to unwieldy graphs with much redundant information. Also, such a representation would depend on the sampling frequency which is often chosen arbitrarily.

**Example 4.2** Consider the following trivariate Gaussian process $X$ with

$$\begin{aligned} X_1(t) &= \alpha_1 Z(t-2) + \beta_{12} X_2(t-1) + \varepsilon_1(t), \\ X_2(t) &= \alpha_2 Z(t-1) + \beta_{23} X_3(t-1) + \varepsilon_2(t), \\ X_3(t) &= \beta_{32} X_2(t-1) + \varepsilon_3(t), \end{aligned}$$

where $Z$ and $\varepsilon_i$, $i = 1, 2, 3$, are independent Gaussian white noise processes with mean 0 and variance $\sigma^2$. The corresponding graph is shown in Figure 1.

In order to be more specific about the properties of such graphs, we need to introduce some notation first.

### 4.2 GRAPH NOTATION

The graphs $G = (V, E)$ used here are so–called mixed graphs that may contain two types of edges directed edges $a \to b$ or $a \leftarrow b$, and (dashed-) undirected[1] edges $a$ --- $b$ for distinct nodes $a, b$. Multiple edges between two nodes are allowed if they are of different type or orientation, i.e. there can be up to three edges between two nodes. Most of the terminology known for directed acyclic graphs can still be applied for these mixed graphs except for the notion of paths. A path in our graphs cannot uniquely be defined by a sequence of nodes as there may be different edges between two nodes. Hence a *path* $\pi$ from $a$ to $b$ is defined as a sequence $\pi = (e_1, \ldots, e_n)$ of *not necessarily distinct* edges $e_i \in E$, such that $e_i$ is an edge between $v_{i-1}$ and $v_i$ for some sequence of *not necessarily distinct* vertices $v_0 = a, v_1, \ldots, v_n = b$. A path $\pi$ in $G$ is called a *directed path* if it is of the form $a \longrightarrow \ldots \longrightarrow b$ or $a \longleftarrow \ldots \longleftarrow b$. Similarly, if $\pi$ consists only of undirected edges, it is called an *undirected path*. Furthermore, a path between vertices $a$ and $b$ is said to be $b$-pointing if it has an arrowhead at the endpoint $b$, that is, $e_n = v_{n-1} \longrightarrow b$. More generally, we call a path a $B$-pointing path if it is $b$-pointing for *some* $b \in B$. Similarly, we call a path between vertices $a$ and $b$ *bi–pointing* if it has an arrowhead at both endpoints, that is, $e_1 = a \longleftarrow v_1$ and $e_n = v_{n-1} \longrightarrow b$. In particular we will make use of the following definition.

---

[1]In contrast to Richardson (2003), we use dashed undirected edges --- instead of bi–directed edges $\leftrightarrow$ as we use directed edges exclusively for indicating direction in time; dashed edges with a similar connotation have been used by Cox and Wermuth (1996).

**Definition 4.3 (Front– and back–door paths)** Let $\pi = (e_1, \ldots, e_n)$ be a path from $a$ to $b$. We say that $\pi$ is a *front-door path* from $a$ to $b$ if $e_1 = a \longrightarrow v_1$. Otherwise we call $\pi$ a *back-door path* from $a$ to $b$.

As in Frydenberg (1990), a node $b$ is said to be an *ancestor* of $a$ if either $b = a$ or there exists a directed path $b \longrightarrow \cdots \longrightarrow a$ in $G$. The set of all ancestors of elements in $A$ is denoted by $an(A)$, which by definition includes $A$ itself. Notice that this differs from Lauritzen (1996). A subset $A$ is called an *ancestral set* if it contains all its ancestors, that is, $an(A) = A$.

For our mixed graphs the concept of separation is based on the following notion of *colliders*. An intermediate vertex $c \in \{v_1, \ldots, v_{m-1}\}$ on a path $\pi$ is said to be an *m–collider* if the edges preceding and succeeding $c$ on the path both have an arrowhead or a dashed tail at $c$ (e.g. $\longrightarrow c \longleftarrow$, $--- c ---$, $--- c \longleftarrow$); otherwise the vertex $c$ is said to be an *m–non–collider* on the path (e.g. $\longrightarrow c \longrightarrow$, $--- c \longrightarrow$, $\longleftarrow c \longrightarrow$). Notice that at least one of the edges that are adjacent to an $m$–non–collider $c$ on a path must be a directed edge with a tail at $c$. Also notice that endpoints are neither colliders nor non–colliders. Furthermore, if the path passes through a vertex $c$ more than once, this vertex may be a $m$–collider as well as a $m$–non–collider depending on its position on the path. With these definitions, a path $\pi$ between vertices $a$ and $b$ is said to be $m$–*connecting* given a set $S$ if

(i) every $m$–non–collider on the path is not in $S$, and
(ii) every $m$–collider on the path is in $S$,

otherwise we say the path is $m$–*blocked* given $S$. Note that the path is also blocked when the same node is a $m$–collider and a $m$–non–collider at different stages on the path because then (i) and (ii) cannot both be satisfied.

**Definition 4.4 ($m$–separation)** Let $G$ be a mixed graph, and let $a, b \in V$ and $S \subseteq V$. If all paths between $a$ and $b$ are $m$–blocked given $S$, then the vertices $a$ and $b$ are said to be $m$–*separated* given $S$. Similarly, if $A$ and $B$ are disjoint subsets of $V$, the sets $A$ and $B$ are said to be $m$–separated given $S$ if, for every pair $a \in A$ and $b \in B$, $a$ and $b$ are $m$–separated given $S$.

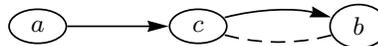

Figure 2: Example of a mixed graph.

**Example 4.5** In the simple graph given in Figure 2 we find that $a$ and $b$ are not $m$–separated by whatever set. This is because there are some paths on which $c$ is a $m$–collider like $a \longrightarrow c --- b$ and some where it is a $m$–non–collider like $a \longrightarrow c \longrightarrow b$. Hence, the latter path is $m$–blocked by $c$ but the former is not.

### 4.3 GRAPHICAL TIME SERIES MODELS

The probabilistic time series model corresponding to a mixed graph has to satisfy the following Markov property which is a generalisation of Eichler (2007) and similar to Eichler (2001).

**Definition 4.6 (Global Markov property)** Let $X$ be a multivariate stationary time series and $G = (V, E)$ be a mixed graph. Then $X$ satisfies the *global Granger–causal Markov property* with respect to $G$ if the following two conditions hold for all disjoint subsets $A$, $B$, and $C$ of $V$.

(i) If every $B$-pointing path between $A$ and $B$ is $m$–blocked given $B \cup C$, then $X_A \not\to X_B \, [X_{A \cup B \cup C}]$.

(ii) If every bi–pointing path between $A$ and $B$ is $m$–blocked given $A \cup B \cup C$ and there is no undirected edge between $A$ and $B$, then $X_A \sim X_B \, [X_{A \cup B \cup C}]$.

Note that in the absence of undirected edges, i.e. in the absence of contemporaneous dependence, the above graphs are the discrete time analogue of 'local independence graphs' for continuous time Markov processes (Didelez, 2006).

**Example 4.7** For the graph in Figure 3 every path between $a$ and $b$ is $b$–pointing. Further we find that $c$ but not $d$ $m$–separates these two nodes. This is because every path has to go through $c$ which is always a $m$–non–collider due to the directed edge to $b$. In contrast, $d$ can be a $m$–collider or $m$–non–collider on different paths and it is hence not enough to condition on $d$. Consequently, $X_a$ is Granger–noncausal for $X_b$ with respect to $X_{\{a,b,c\}}$ but not with respect to $X_{\{a,b,d\}}$. In contrast, $X_b \not\to X_a \, [X_{\{a,b,c\}}]$ as well as $X_b \not\to X_a \, [X_{\{a,b,d\}}]$ because on every $a$–pointing path $c$ as well as $d$ are always $m$–non–colliders. For the same reason $X_a \sim X_b \, [X_{\{a,b,c\}}]$ as well as $X_a \sim X_b \, [X_{\{a,b,d\}}]$.

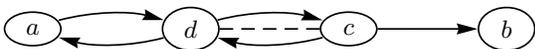

Figure 3: Illustration of global Markov property.

In particular cases we can read off a time series graph when Granger–noncausality holds at all horizons. These are characterised as follows.

**Theorem 4.8** *Suppose that $X$ satisfies the global Granger–causal Markov property with respect to a mixed graph $G$. Let $A, B, C$ be disjoint subsets of $V$. Then if every $an(B)$-pointing path between $A$ and $an(B)$ is $m$–blocked given $B \cup C$, $X_A$ is Granger–noncausal at all horizons with respect to $X_{A \cup B \cup C}$ $(X_A \overset{\infty}{\not\to} X_B[X_{A \cup B \cup C}])$*

**Example 4.9** In Figure 3 only $X_b$ is Granger–noncausal at all horizons for the other components. In particular, as $an(b) = \{a, b, c, d\}$ we have that $X_a$ is not Granger–noncausal at all horizons for $X_b$ given any conditioning set.

### 4.4 BACK–DOOR CRITERION

So far, the above graphs are just representing conditional (in)dependencies between past and present of time series. In order to make causal statements we need to include interventions; the following lemma provides the key in linking such interventions to the graphical representation of time series.

**Lemma 4.10** *Suppose that $X$ satisfies the global Granger–causal Markov property with respect to a mixed graph $G$ (under $\sigma_a(t) = \emptyset$, i.e. in the observational regime) and that Assumptions 2.3 hold. Let $a, b \in S \subseteq V$.*
*If every $an(b)$-pointing back-door path between $a$ and $an(b)$ is $m$–blocked given $S$, then*

$$X_b(t+h) \perp\!\!\!\perp \sigma_a(t) | \overline{X}_S(t)$$

*for all $h \in \mathbb{Z}$, $t \in \mathbb{Z}$.*

**Proof:** In order to see this consider an auxiliary graph $\tilde{G}$ which is an augmented version of $G$ in that it includes an additional node $\tilde{a}$ for a time series $\{F_{\tilde{a}}(t), t \in \mathbb{Z}\}$. This time series indicates when an intervention in $X_a$ takes place. Let $t^*$ be the potential time of the intervention, then $F_{\tilde{a}}$ is defined as $F_{\tilde{a}}(t) = \emptyset$ for $t \neq (t^* - 1)$ and $F_{\tilde{a}}(t^* - 1) = \sigma_a(t^*)$. Note that as it must be decided before $t^*$ whether or not an intervention in $X_a(t^*)$ takes place and $\overline{X}(t^* - 1) \perp\!\!\!\perp \sigma_a(t^*)$ (by Assumptions 2.3) we can assume without loss of generality that the value of $\sigma_a(t^*)$ is known at $t^* - 1$. The node $\tilde{a}$ is included in $\tilde{G}$ with a single directed edge into the node $a$ reflecting Assumptions 2.3. For the same reason the conditional distributions of $X_a(t)$ given the past remain unchanged for $t \neq t^*$ and are determined by Definition 2.1 for $t = t^*$. As every $an(b)$-pointing back-door path between $a$ and $an(b)$ is $m$–blocked given $S$ and $\tilde{a}$ has a directed edge only into $a$, we know that every $an(b)$-pointing back-door path between $\tilde{a}$ and $an(b)$ is $m$–blocked given $S$. Applying Theorem 4.8 yields that $F_{\tilde{a}}$ is noncausal for $X_b$ at all horizons given $S$, so in particular $X_b(t+h) \perp\!\!\!\perp \sigma_a(t) | \overline{X}_S(t)$.

Lemma 4.10 gives sufficient conditions so that the conditional distributions of $X_b(t+h)$ given $\overline{X}_S(t)$ under an interventional and the observational regime are the same. Hence, we can use data from an observational study to predict effects of interventions if the processes in $S$ have been observed. The following is immediately obvious.

**Theorem 4.11 (Back-door criterion)** *The assumptions of Theorem 3.1 are satisfied if all $an(b)$-pointing back-door paths between $a$ and $an(b)$ are $m$–blocked given $S$.*

**Example 4.12** Assume we want to assess the causal effect of $X_a$ on $X_b$ in Figure 3. We know that $X_a$ is not Granger–noncausal for $X_b$ unless we condition on $X_c$ and in particular we know that $X_a$ is not Granger–noncausal for $X_b$ at all horizons. This raises the question of $X_a$'s actual causal effect on $X_b$. To check identifiability of this effect we consider every back–door path from $a$ to $an(b) = \{a, b, c, d\}$. All these paths will start with $a \longleftarrow d$ hence $d$ is always a $m$–non–collider. Therefore we choose $S = \{a, b, d\}$ and can apply the above theorem. If $X_d$ is a latent unobservable process then the causal effect cannot be identified. Note that conditioning on $c$ is not required and not enough.

To illustrate the actual calculation of the ACE using the back–door criterion consider again Example 4.2.

**Example 4.2 ctd.** *Suppose that we are interested in the effect of an intervention $s$ setting $X_3(t)$ to $x_3^*$ on $X_1(t+2)$. Simple calculations show that $X$ has the autoregressive representation*

$$X_1(t) = \left(\frac{\alpha_1 \alpha_2}{1 + \alpha_2^2} + \beta_{12}\right) X_2(t-1)$$
$$- \frac{\alpha_1 \alpha_2 \beta_{23}}{1 + \alpha_2^2} X_3(t-2) + \tilde{\varepsilon}_1(t), \quad (4)$$
$$X_2(t) = \beta_{23} X_3(t-1) + \tilde{\varepsilon}_2(t),$$
$$X_3(t) = \beta_{32} X_2(t-1) + \tilde{\varepsilon}_3(t)$$

*where $\tilde{\varepsilon}_i$, $i = 1, 2, 3$, are again independent zero mean Gaussian white noise processes and independent of the other $X$ components. Hence, from the full model, we immediately obtain that*

$$\mathbb{E}_s X_1(t+2) = \beta_{12} \beta_{23} x_3^*.$$

Now suppose that only $X$ has been observed and the process $Z$ takes the role of an unobserved variable. To apply Theorem 3.1, we note that in Figure 1 every pointing back–door path between 3 and some other node $v$ is bi–pointing starting with the edge $3 \longleftarrow 2$ and hence is m–blocked given $S = \{1,2,3\}$ because 2 is a noncollider. Thus, $S$ identifies the effect of $X_3(t)$ on $X_1(t+2)$ and the average causal effect can be obtained from the autoregressive representation of $X$ in (4) by

$$\mathbb{E}_s X_1(t+2) = \phi_{13}^{(2)}(1) = \phi_{12}(1)\phi_{23}(1) + \phi_{13}(2)$$
$$= \beta_{12}\beta_{23}x_3^*.$$

### 4.5 FRONT–DOOR CRITERION

Due to space limitation we do not go into the technical details of the front–door criterion but rather give an idea of its use with an example. The following theorem states graphical conditions, that are sufficient to identify the causal effect and are different from the back–door conditions.

**Theorem 4.13** *Suppose that $X$ satisfies the global Granger–causal Markov property with respect to a mixed graph $G$ (under $\sigma_a(t) = \emptyset$). Further, let $a,b \in S \subset V$ and let $C = S\setminus\{a,b\}$. If the following graphical conditions hold the causal effect of $X_a(t)$ on $X_b(t+h)$ can be identified.*

1. *Every directed path from $a$ to $b$ is m–blocked given $C$.*
2. *Every an$(C)$-pointing back-door path between $a$ and an$(C)$ is m–blocked given $S$.*
3. *Every an$(b)$-pointing back-door path between $C$ and an$(b)$ is m–blocked given $S$.*

**Proof:** Eichler and Didelez (2007).

Notice that the second and third conditions are the same as in Theorem 3.1 and ensure that the causal effects of $X_a$ on $X_C$ as well as of $X_C$ on $X_b$ can be identified.

**Example 4.14** For an illustration, consider the trivariate process $X$ given by

$$X_1(t) = \alpha Z(t-2) + \beta_{12} X_2(t-1) + \varepsilon_1(t),$$
$$X_2(t) = \beta_{23} X_3(t-1) + \varepsilon_2(t),$$
$$X_3(t) = \varepsilon_3(t),$$

where $Z$ and $\varepsilon_i$, $i = 1,2,3$, are Gaussian white noise processes with mean 0, variance $\sigma^2$, and

$$\text{corr}\,(Z(t), \varepsilon_3(t)) = \rho,$$

while all other correlations between $Z$ and $\varepsilon_i$, $i = 1,2$, are 0. Again, the process $Z$ takes the role of an unobserved variable. The path diagram associated with $X$

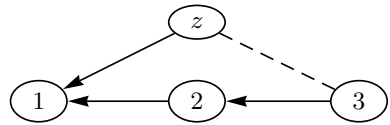

Figure 4: Path diagram associated with the processes $X$ and $Z$ in Example 4.14.

and $Z$ is shown in Figure 4. Simple calculations show that $X$ has the autoregressive representation

$$X_1(t) = \beta_{12} X_2(t-1) + \alpha\rho X_3(t-2) + \tilde{\varepsilon}_1(t),$$
$$X_2(t) = \beta_{23} X_3(t-1) + \tilde{\varepsilon}_2(t), \qquad (5)$$
$$X_3(t) = \tilde{\varepsilon}_3(t),$$

where $\tilde{\varepsilon}_i$, $i = 1,2,3$, are again independent zero mean Gaussian white noise processes, and independent of the $X$ variables in their respective equations.

Now suppose that we are interested in the average causal effect of $X_3(t)$ on $X_1(t+2)$. Since $3 \dashl\dashl\dashl z \rightarrow 1$ constitutes an an(1)-pointing back-door path that is m–connecting given $S = \{1,2,3\}$, we cannot apply the back-door criterion. On the other hand, the graphical conditions for the front-door criterion are easily verified: First, the only directed path from 3 to 1 is m–blocked by node 2. Second, noting an$(2) = \{2,3\}$, every $\{2,3\}$-pointing back-door path between 3 and 2 must end with the edge $3 \longrightarrow 2$ and thus is m–blocked by $3 \in S$. Finally, every an(1)-pointing back-door path between 2 and 1 starts with the edge $2 \longleftarrow 3$ and thus is also m–blocked given $3 \in S$. Therefore, by Theorem 4.13, the average causal effect of setting $X_3(t) = x_3^*$ on $X_1(t+2)$ is identified and can, in the present case, be calculated as

$$\mathbb{E}_s X_1(t+2) = \phi_{12}(1)\phi_{23}(1)\,x_3^* = \beta_{12}\beta_{23}\,x_3^*,$$

where $\phi_{ab}(1)$ are the coefficients of the autoregressive representation in (5).

## 5 CONCLUSION

In conclusion, let us summarise the relation between Granger–(non)causality and causal effects in terms of interventions. It can easily be seen, by a similar reasoning as for Lemma 4.10, that if $a \rightarrow b \notin E$ then $X_a(t)$ has also no causal effect on $X_b(t+1)$ and that if $a \notin \text{an}(b)$ then it has no causal effect on $X_b(t+h)$, $h \in \mathbb{N}$, in the sense that $\text{ACE}_s = 0$. Hence, as mentioned in the introduction, we can regard the directed edges, i.e. Granger causal relations, as potentially causal. However, we emphasise that in addition we require Assumptions 2.3 to be satisfied, which means we need to be able to actually carry out an in-tervention in $X_a(t)$ and the set $V$ of considered components of the multivariate time series must be 'rich' enough to satisfy these assumptions. Granger (1969) has not considered the question of whether an intervention is at all possible, but he has considered the latter by requiring that $X_V$ consists of all 'relevant' processes. To our knowledge, White (2006) is one of the few who addresses the effect of interventions in time series in a similar sense as considered here. The back– and front–door criteria given in the present paper then ensure that this causal effect of an intervention can be estimated from the observable data.

An open question is the use of the proposed graphical representation of multivariate time series based on Granger–noncausality to address *multiple* interventions at sequential points in time (such as considered in Robins, 1986; Pearl and Robins, 1995; Dawid and Didelez, 2005), which might be of particular interest in a dynamic context such as time series. This will be addressed in more detail in Eichler and Didelez (2007).

## Acknowledgements

The research of Michael Eichler was partly financially supported by the Research Training Network DYNSTOCH of the European Community's Human Potential Programme (contract HPRN-CT-2000-00100). Vanessa Didelez acknowledges support from the Centre for Advanced Study at the Norwegian Academy of Science and Letters.